\documentclass[manuscript]{acmart}
\AtBeginDocument{%
  \providecommand\BibTeX{{%
    \normalfont B\kern-0.5em{\scshape i\kern-0.25em b}\kern-0.8em\TeX}}}

\copyrightyear{2022} 
\acmYear{2022} 
\setcopyright{acmcopyright}\acmConference[CHI '22]{CHI Conference on Human Factors in Computing Systems}{April 29-May 5, 2022}{New Orleans, LA, USA}
\acmBooktitle{CHI Conference on Human Factors in Computing Systems (CHI '22), April 29-May 5, 2022, New Orleans, LA, USA}
\acmPrice{15.00}
\acmDOI{10.1145/3491102.3502063}
\acmISBN{978-1-4503-9157-3/22/04}

\usepackage{xcolor}
\usepackage{subcaption}
\usepackage{microtype}
\usepackage{array}
\usepackage{enumitem}
\usepackage{colortbl}
\usepackage{graphicx}

\newcolumntype{\$}{>{\global\let\currentrowstyle\relax}}
\newcolumntype{^}{>{\currentrowstyle}}

\def\markup{0}
\if\markup1
\newcommand{\rv}[1]{{\leavevmode\color{blue}#1}}
\else
\newcommand{\rv}[1]{#1}
\newcommand{\st}[1]{}
\fi

\begin{document}

% \title{``I Choose Assistive Devices That Save My Face''}
% \subtitle{A Study of Practices and Challenges with Assistive Technology Use Conducted in China}
\title{``I Shake The Package To Check If It's Mine''}
\subtitle{A Study of Package Fetching Practices and Challenges of Blind and Low Vision People in China}

\author{Wentao Lei}
\affiliation{%
 % \institution{School of Information}
  \institution{Rochester Institute of Technology}
 % \city{New York}
  \country{USA}
}
\email{wl6506@rit.edu}

\author{Mingming Fan}
\authornote{Corresponding Author}
\affiliation{%
  \institution{Computational Media and Arts Thrust,}
  \institution{The Hong Kong University of Science and Technology (Guangzhou)}
  \city{Guangzhou}
  \country{China}
}
\affiliation{%
  \institution{Division of Integrative Systems and Design,} 
  \institution{Department of Computer Science and Engineering,}
  \institution{The Hong Kong University of Science and Technology}
  \city{Hong Kong SAR}
  \country{China}
}
\email{mingmingfan@ust.hk}

\author{Juliann Thang}
\affiliation{%
  %\institution{School of Information}
  \institution{Rochester Institute of Technology}
 % \city{New York}
  \country{USA}
}
\email{jxt1776@rit.edu}

\renewcommand{\shortauthors}{Lei, Fan and Thang}

\begin{abstract}
With about 230 million packages delivered per day in 2020, fetching packages has become a routine for many city dwellers in China. When fetching packages, people usually need to go to collection sites of their apartment complexes or a KuaiDiGui, an increasingly popular type of self-service package pickup machine. However, little is known whether such processes are accessible to blind and low vision (BLV) city dwellers. We interviewed BLV people (N=20) living in a large metropolitan area in China to understand their practices and challenges of fetching packages. Our findings show that participants encountered difficulties in finding the collection site and localizing and recognizing their packages. When fetching packages from KuaiDiGuis, they had difficulty in identifying the correct KuaiDiGui, interacting with its touch screen, navigating the complex on-screen workflow, and opening the target compartment. We discuss design considerations to make the package fetching process more accessible to the BLV community.
  
%   proposed potential solutions and their limitations for different challenges. In this way, future engineers and researchers can better understand the situations of BLV people receiving packages, and take the accessibility issues into consideration when developing delivery products .
  
%   And due to the lack of screen reader for self-service package pickup machine, the BLV people failed to fetc
  
%   In the study, we found that BLV people  package localization and recognition, usage of KuaiDiGui ( a self-pickup package machine), and thinking of BLV identity are emphatically mentioned by the BLV participants because of lacking corresponding accessibility technology. 
  
%   Based on the findings,  we further proposed several potential solutions and their limitations for different challenges. In this way, future engineers and researchers can better understand the situations of BLV people receiving packages, and take the accessibility issues into consideration when developing delivery products .

\end{abstract}

\begin{CCSXML}
<ccs2012>
<concept>
<concept_id>10003120.10011738.10011773</concept_id>
<concept_desc>Human-centered computing~Empirical studies in accessibility</concept_desc>
<concept_significance>500</concept_significance>
</concept>
</ccs2012>
\end{CCSXML}

\ccsdesc[500]{Human-centered computing~Empirical studies in accessibility}

\keywords{Package delivery, KuaiDiGui, Blind and low vision, People with vision impairments, Qualitative study, Interview, China, Accessibility}

\maketitle

\section{Introduction}
More than 2.2 billion people around the world have different levels of vision impairment including around 40 million blind people \cite{WHO2019RV}. China has more than 17 million blind and visually impaired (BLV) people including 5 million blind people \cite{ChinaDisablityReport}. 
In the meantime, around 782 million people performed online shopping on domestic e-commerce platforms in China and more 230 million packages were delivered from these platforms to customers in 2020 alone\cite{internetdevelopment}. This suggests that about one-third of Chinese people had experience fetching packages every day in 2020 \cite{Index2020}. As an indispensable and important part of the society, BLV people in China are expected to perform online shopping and package fetching in their daily lives. %Online shopping and fetching delivery packages become inseparable activities for the blind and visually impaired people's daily lives. 

For the ``Last Mile Delivery'', delivery services in Europe or America areas usually send packages to the recipient’s house/apartment door directly, which result in relatively less confusion; however, in China, many city dwellers live in a residential apartment complex \cite{Census2021, HousinginChina2020,ShanghaiResidence}, which is the most common living environment for BLV people \cite{ShanghaiResidence}. Delivery services usually send the packages of the residents in many apartment buildings to a collective site first then notify the recipients to fetch their packages by themselves. Consequently, fetching a package is a highly visually-demanding task because people need to search through piles of packages, read, and discern the printed labels on packages. 
Thus, this process can be particularly challenging for BLV people. However, it remains unknown \textit{whether and to what extent package fetching process is accessible to BLV people and the challenges they are facing.} 

What's more, in recent years, self-pickup delivery methods such as \textit{self-service package pickup machine}, have been widely adopted by many cities in China. Such pickup machines, known as KuaiDiGui, often have a touchscreen and require inputting a passcode to open a compartment of the machine, which typically contains a large grid of compartments (see Figure~\ref{fig:KDG}). However, little is known about \textit{whether and to what extent such self-service package pickup machine is accessible to BLV people and the challenges they may face.}

% However, this visually-demanding process can be challenging for BLV city dwellers. In particular after COVID, it has been becoming a norm to shop online, and fetching packages of online shopping is an increasingly urgent need for the BLV community.  

% When localizing packages, there would be various types of objects in the environment, which make it hard for BLV people to find the position of packages. Also, when multiple packages belonging to different people piling at the same place, finding their package and identify the recipient information on the packages are significantly difficult for BLV people. In terms of self-service package pickup machine, users need to input pickup code in the machine systems to accomplish pickup process. However, due to the lack of accessible technology (e.g., screen reader), many BLV people fail to do that because they cannot manipulate the touch screen by clicking randomly.

% Previous researchers have put lots of efforts on how to improve the information system accessibility for blind and visually impaired people. Some researchers developed applications using captured pictures or videos to achieve object recognition and indoor navigation via machine learning technology like computer version and deep network algorithm. However, there is little research investigating the delivery package fetching process for BLV people.

Motivated by this need, we sought to answer the following research questions (RQs):

\begin{itemize}%[leftmargin=*]
\item {RQ1}: What are the current practices and challenges of fetching delivery package for BLV people?
\item {RQ2}: What are the design considerations for improving the accessibility of package fetching for BLV people?
\end{itemize}

To answer the RQs, we conducted a semi-structured interview study with 20 BLV participants to understand their practices and challenges of fetching packages. 
%how they fetch the delivery packages, and the challenges that they encountered during the process. We sought to answer the following three research questions (RQS):
% \newline1) what are the current practices of fetching delivery package for BLV people? \newline2) What are the challenges the BLV population would encounter while performing the package fetching? \newline3) What changes could we propose to improve the accessibility performance of the package delivery process for BLV population?
Our findings show that participants tended to use their limbs or white canes to poke and explore a package collection site to locate packages; however, this approach did not work well for localizing small-size packages. To recognize the identity of a package, participants relied on either their residual vision or tactile feedback of the package's surface to first locate the printed label area and then use their phone's Optical Character Recognition (OCR) function or take a picture and zoom in on the picture to recognize the content on the label. Participants experienced difficulties in all of these steps. Similarly, participants encountered many challenges with KuaiDiGui (i.e., the self-service package pickup machine), in particular, locating the right KuaiDiGui from similar KuaiDiGuis in the vicinity, manipulating the touchscreen to open the target compartment, and locating the open compartment. Furthermore, we uncovered issues with seeking help, including deciding when and whom to seek help from and self-impression management. Finally, based on our findings, we present design considerations to make the package fetching process more accessible to BLV people. To our knowledge, this is the first study to explore BLV people's practices and challenges of fetching packages in light of the increasingly popular online shopping and package delivery services. In summation, we make the following contributions: 

\begin{itemize}%[leftmargin=*]
\item We show the practices and challenges of how BLV people fetch packages, in light of increasingly pervasive online shopping, from package collection sites and self-service package pickup machines, two common approaches to receiving packages for online shopping;
\item We present implications of the findings and offer design considerations to improve the accessibility of package fetching process for BLV people.
\end{itemize}

%on package label or zoom in the label image on their mobile phone. This process is very long and complicated because 

%it’s hard for BLV people to localize and frame the label well with their mobile phone cameras. For the use of self-service package pickup machine, we found that it is totally inaccessible for blind users because of the lacking accessible technologies for the pickup machine touch screen, and it’s challenging for BLV people to localize the opening compartments on the machines. In terms of help seeking and thinking of BLV identity, we found that most BLV people do not want to applied too much on other sighted people, however, when dealing with problems they cannot solve independently indeed, they would turn to others for helps and have their own preferences for target helpers.Based on these findings, the study discussed the design and research opportunities to improve the pacakge fecthing accessibility for BLV population.
\section{BACKGROUND AND RELATED WORK}

% \subsection{Package delivery}

\st{The courier, express, and parcel (CEP) market reached 342.61 billion Euros in 2019. In China the CEP industry business volume reached 83.36 billion cases in 2020, and the annual revenue reached 879.54 billion yuan taking more than 60 $\%$ of the global CEP volume . In 2020, more than 230 million packages were handled by Chinese express companies, and 460 million Chinese people get package services every day. In this way, the CEP industry has become an inseparable part of Chinese people's daily lives.}

% \subsection{Package delivery}

% The courier, express, and parcel (CEP) market reached 342.61 billion Euros in 2019 \cite{Statista2020}. In China the CEP industry business volume reached 83.36 billion cases in 2020, and the annual revenue reached 879.54 billion yuan taking more than 60 $\%$ of the global CEP volume \cite{CEPbulletin2020, Index2020}. In 2020, more than 230 million packages were handled by Chinese express companies, and 460 million Chinese people get package services every day \cite{Index2020}. In this way, the CEP industry has become an inseparable part of Chinese people's daily lives.

\subsection{Delivery Methods in China}
Different from North America or European countries, China is a crowded country with 1.43 billion people \cite{Census2021, HousinginChina2020} and has a large number of high density residential communities with lots of apartment complexes, which is also the most common living environment for BLV people in China \cite{ShanghaiResidence, HousinginChina2020}. To better serve the customers living in crowded communities, the CEP companies introduced various delivery methods, including \textit{door-to-door delivery}, \textit{package pickup site}, and most recently the \textit{self-service package pickup machine}.

\subsubsection{Door-to-door Delivery}
The delivery workers send the packages to the customer's house door directly. One situation is when the recipient has physical contact with the delivery worker where face-to-face communication (e.g., check the packages) might occur during the delivery process. The second situation is there is no physical contact during the delivery. The delivery worker leaves the packages at a certain location outside the door, which happens a lot during the COVID-19 pandemic. For the second situation, BLV people often need to localize and recognize the packages by themselves.

\subsubsection{Collective Pickup Site} 
The delivery workers send the packages for residents from surrounding areas to a fixed collective site (e.g., stores, community offices, and package stations). The recipients could come to the collective sites for pickup by communicating with the on-site staff. In 2020, there were 114,000 pick-up sites in China \cite{Index2020}.

\begin{figure}[htb!]
  \centering
  \includegraphics[width=\linewidth]{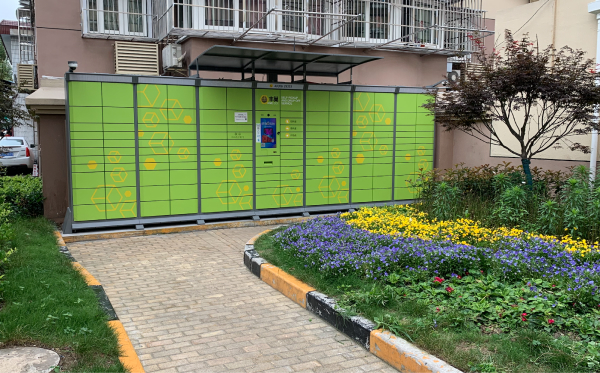}
  \caption{FengChao KuaiDiGui outside the apartment buildings}
  \label{fig:KDG}
  \Description{A picture that shows a Fengchao KuaiDiGui standing beside a pathway inside a community.}
\end{figure}

\subsubsection{KuaiDiGui} This refers to the self-service package pickup machine for storing and picking up packages, which is similar to the Amazon lockers in USA. As seen in Figure \ref{fig:KDG}, the KuaiDiGui usually appears as a large iron cabinet with many compartments for storing packages. A standard Fengchao KuaiDiGui has 84 compartments and a touchscreen for delivery workers and other users to operate the system  \cite{Fengchao}. In 2020, more than 400,000 sets of KuaiDiGui were deployed throughout China \cite{Index2020}.

\begin{figure*}[htb!]
  \centering
  \includegraphics[width=\linewidth]{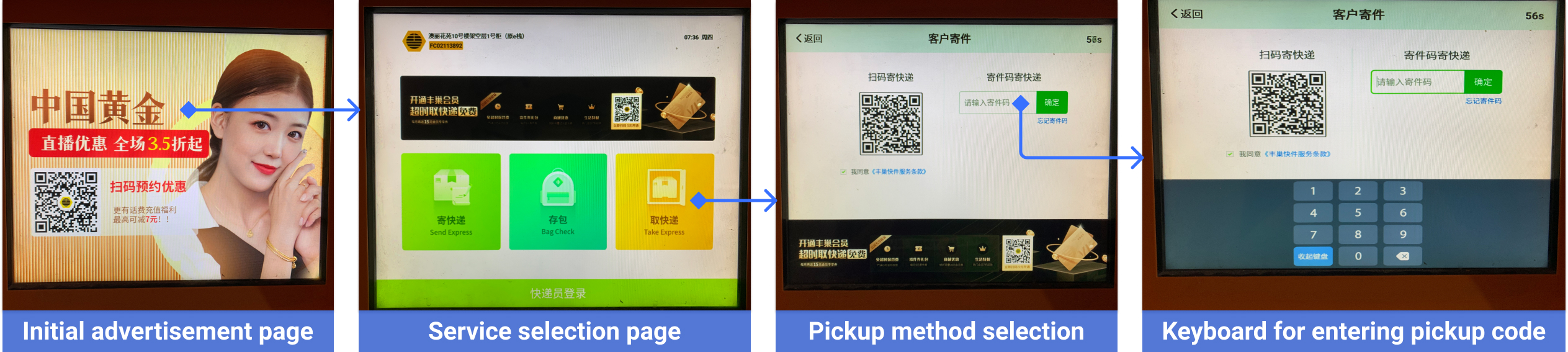}
  \caption{Task flow of Fengchao KuaiDiGui touchscreen}
  \label{fig:FCinterface}
  \Description{Four images that show the task flow of using the Fengchao KuaiDiGui’s touch screen to pickup, including initial advertisement page, service selection page, pickup method selection page, and keyboard for entering pickup code page.}
\end{figure*}

After the delivery workers put packages into the compartments, the KuaiDiGui system will automatically send a notification message including pickup code, KuaiDiGui's location and other information to the recipients. After recipients find the KuaiDiGui's location, they need to operate the touchscreen for pickup. Take Fengchao KuaiDiGui as an example, as seen in Figure \ref{fig:FCinterface}, the user needs to first click on the initial advertisement page and then the system will jump to the service selection page; by clicking the "Take Express" button, the system will provide two pickup methods, users can either use their smart phones to scan the QR code or enter an 8 digit pickup code on the numeric keyboard. After these operations, the system will check whether there is a package for the user. If there is, a compartment door will open accordingly and the user can take the package out from it. Currently, only a few KuaiDiguis have voice feedback when clicking on the touchscreen elements, and no KuaiDigui has a screen reader feature.

\subsubsection{Outdoor Package Pile} This refers to putting packages in piles or on a shelf in a certain place without delivery workers on-site andthe recipients need to go through all the packages one by one themselves. During the COVID-19 pandemic, some communities in China prohibited delivery workers from entering communities for delivery, resulting in many packages being piled outside of the residence buildings. 

\subsection{Accessibility Research for BLV Population}
To our knowledge, there is no previous research investigating the package fetching process for the BLV population. However, many previous studies have focused on how to improve the accessibility level of various activities for BLV population, including indoor/outdoor navigation, usage of mobile devices, object identification, and etc., which were related to each step of the package fetching process.

\subsubsection{Navigation in the City/Building}
\st{To fetch packages at the collective package sites, BLV people need to leave their houses, set the routes, travel through the community, and identify if they reach the right destination. For indoor navigation, previous research} \rv{Previous research has} proposed multiple approaches such as sensor models, augmented reality, and information label systems to improve travel accessibility for the BLV population \cite{yuhang2019AR, yuhang2019head-mounted, Andreas2003Indoor, gleason2016vizmap, herskoviz2020ARaccessible}. Zhao et al. explored augmented reality (AR) technologies on projection platforms, smart glasses, and head mounted display devices to enhance real-world vision through projection and image processing to help low vision people better navigate an indoor environment \cite{yuhang2019AR, yuhang2019head-mounted}. In terms of city navigation, some previous research worked on \rv{how BLV people feel about }the accessibility of public transportation which included bus taking, train travelling, and ride-hailing\st{, and how BLV people feel about the public transportation process.}\cite{Shiri2011pubilictransit,Megan2014Stopinfo,vaishnav2018ridehailing,Robin2019Trustridesharing}.

\st{For our research, we investigated} \rv{To fetch packages at collective package sites, BLV people need to leave their houses, set routes, travel through the community, and identify if they have reached the right destination. Our research investigates }the process of how BLV people \rv{navigate outside their residence} to collective package sites\rv{. Specifically, we focus on} what technologies and approaches they use\st{d for traveling and navigation process.} \rv{to reach their package destination, the issues they faced while travelling, and how they feel the process could be improved.}

\subsubsection{Usage of Touchscreen Devices}
BLV people can perform tasks such as online shopping, tracking logistical information of packages, and receiving package messages on their mobile devices (e.g., mobile phones, computers, smart watches). In order to receive information from devices and to perform operations, screen readers play an important role on mobile phones. Previous researchers have conducted studies to understand the general accessibility and usability issues of screen readers on mobile phones and how to make visual content (e.g. texts, images, UI elements and components) more accessible for screen readers \cite{Barbara2012Voiceover, Morris2018Visualcontent, lucas2016readercontents, anne2018buttonlabel, anne2017epidemiology, xiaoyi2021metadata, peng2021sayitall}. \rv{For touchscreen accessibility, researchers have investigated how screen size and key size could affect BLV people’s interaction with touchscreens} \cite{tiago2015touchscreen, rodrigues2016keysize}. \rv{Specifically for kiosks, researchers have investigated the efficiency of different touchscreen gestures for blind users as well as utilizing multiple sensory resources for input} \cite{SandnesGestures, KioskPrototype2014}. \rv{Our study will focus on how BLV people feel and utilize different touchscreen devices in their package fetching process.}

\rv{During the package fetching process, sometimes BLV people need to interact with the interface of KuaiDiGui, which utilizes a touchscreen, often lacks voice feedback, and has no screen reader feature. The lack of accessible features is also very common for other kiosk systems (e.g. ATMs, voting machines, self-service vending machines). There are currently no unified kiosk accessibility guidelines, though, researchers have proposed a set} \cite{KioskGuide2019} \rv{to mimic the Web Content Accessibility Guidelines (WCAG), the golden standard for Web accessibility. Other researchers have proposed multiple approaches to improve the accessibility of kiosk interfaces. Bidarra et al. designed an interactive kiosk prototype with accessible features (e.g. screen reader, text magnifier, tactile mouse) for low vision and elderly populations} \cite{KioskPrototype2014,KioskPrototype2013}. \rv{Guo et al. presented multiple applications (e.g. Vizlens,  StateLens,  ApplianceReader) that would send interface images to a crowd system, label the interfaces via crowd workers, then describe the interface elements beneath users’ fingers via mobile phone screen reader} \cite{anhong2015appliancereader, Anhong2016vizlens, anhong2017realworldinterfaces, anhong2019statelens}. \rv{Previous kiosk-related research mainly focused on improving the accessibility of screen interfaces. However, to fetch packages from KuaiDiGuis, users not only have to operate the screen interface and enter the pickup code correctly, but they also need to find the open compartment and take out their packages, creating additional challenges.}

% In our study, we investigated what challenges the BLV people encountered in each step of interacting with KuaiDiGuis and discussed the possible intervention to improve the accessibility.

% Other approaches were \st{also} presented to support \st{information} reading and writing \rv{information }on mobile devices\st{, including} \rv{such as} braille \st{reading and writing} \cite{nicolau2013braille, daniel2018braillerinput}\st{,} and 3D-printed tactile interfaces for operations \cite{zhang20183Dtactilem, anhong2019statelens}.

\rv{Although previous research has studied the accessibility issues of touchscreen devices for BLV people, to our knowledge, no study has focused on the KuaiDiGui system.} Since multiple interfaces (e.g. mobile phones, KuaiDiGuis) are involved during the package fetching process and the accessibility situation of these interfaces stays uncertain, our study aims to investigate\st{ whether the involved interfaces are accessible enough for BLV people and}\rv{how BLV people feel about the accessibility of these interfaces when fetching a package}.

\subsubsection{Recognize Objects}
Localizing and recognizing the correct packages are critical steps in the package fetching process. Some researchers have developed applications using captured pictures or videos to achieve object recognition via computer vision and deep network algorithms \cite{yuhang2018facerecognition, yu2013realtime, dragan2020ReCog,rodrigues2017hintme, PeneuFetch}. Previous research has worked on various approaches to \st{achieve identification of} \rv{identify} text-based information (e.g., currency bills, merchant characters and clothes labels) \cite{ Remi2019Object, Noboru2013Character, Michele2013Tag}. A key pain point for BLV people in using object recognition technology is \st{hard to aim}\rv{the difficulty with aiming at} the right targets and tak\st{e}\rv{ing} high quality image\rv{s} for recognition. Ahmetovic et al. designed a camera-aiming guidance system to help blind users adjust camera angles \cite{dragan2020ReCog}\st{, and} Zhong et al. introduced an approach by extracting high-quality photos from continuous camera \st{video}\rv{footage} and using \rv{a }cloud-based search engine to match the best result\rv{s to the images} \cite{yu2013realtime}. \st{so BLV people no longer need to aim the targets precisely}\rv{These approaches allowed BLV people to no longer need to precisely aim at targets.}

In our study, we investigate\st{d how} BLV people's practices of localizing and recognizing their packages\st{, and what}\rv{along with the} challenges they \st{would} encounter\rv{ed} when performing these activities. \rv{ Specifically, we investigate how image capturing for text on package labels plays a key role in the package recognition process.}

\begin{table*}[t]
\caption{Participants' demographic information.} % title of Table
\centering % used for centering table

\small
\begin{tabular}{|p{0.2cm}|p{0.4cm}|p{0.8cm}|p{3.4cm}|p{2cm}|p{2cm}|p{2cm}|p{3cm}|} % centered columns (8 columns)

\hline %inserts double horizontal lines
ID & Age & Gender & Visual impairment level & Color perception & Contrast sensitivity & Congenital visually impaired & Occupation \\ [0.5ex] % inserts table
%heading

\hline % inserts single horizontal line
1 & 34 & M& Totally blind & No & No & Yes & Massager \\
 \hline
2 & 24 & F& Totally blind & No & Yes & Yes  & Massager \\ \hline
3  & 30 & M & Totally blind & No & No & Yes & Massager \\ \hline

4 & 23 & F & Blind with residual vision& Yes & No & Yes & Data annotator \\
 \hline

5 & 31 & M & Blind with residual vision& Yes & No & Yes & Massager \\
 \hline

6 & 34 & F & Blind with residual vision& Yes & No & Yes & Receptionist \\
 \hline

7 & 34 & M & Blind with residual vision& Yes & Yes & Yes & Company staff \\
 \hline

8 & 46 & M & Totally blind& No & No & Yes & Unemployed \\
 \hline

9 & 30 & M & Low vision& Yes & Yes & Yes & Freelance \\
 \hline

10 & 33 & F & Blind with residual vision& Yes & Yes & Yes & Company staff \\
 \hline

11 & 38 & F & Totally blind & No & No & Yes & Part-time worker \\
 \hline

12 & 39 & M & Blind with residual vision& Yes & Yes & No & Volunteer organization staff \\
 \hline

13 & 18 & M & Low vision& Yes & Yes & Yes & High school student \\
 \hline

14 & 36 & M & Blind with residual vision& Yes & Yes & No & Affiliated staff \\
 \hline

15 & 32 & M & Blind with residual vision& Yes & Yes & No & Affiliated staff\\
 \hline

16 & 35 & M & Blind with residual,retinitis pigmentosa vision& Yes & Yes & Yes & Part-time worker \\
 \hline

17 & 35 & F & Totally blind & No & Yes & Yes & Food appreciator\\
 \hline

18 & 29 & M & Totally blind& No & No & No & Freelance\\
 \hline

19 & 37 & M & Totally blind& No & No & No & Volunteer organization staff \\
 \hline

20 & 32 & M & Blind with residual vision & Yes & Yes & No & Volunteer organization staff \\
 \hline

\end{tabular}

\label{table:demographic} % is used to refer this table in the text
\end{table*}

\section{METHOD}
\label{Method}
We conducted an IRB-approved interview study with BLV people to gather information of their practices and challenges with fetching delivery packages. We followed local COVID social distancing regulations when conducting the interviews. All interviews lasted around 45 - 90 minutes and were audio recorded. The records were transcribed verbatim for the thematic coding.

\subsection{Participants}
To deeply investigate the practices and challenges with fetching deliveries of visually impaired population. All participants were required to have experiences of fetching deliveries and have certain level of visual impairments including blind and low vision. 
We posted advertisements in local BLV community and volunteer groups. In the end, we recruited 20 participants with various level of visual impairment. Table~\ref{table:demographic} shows their demographic information. 
All participants were authorized as being in the visually impaired population by China Disabled Persons’ Federation. Participants were between 18 and 46 years old, including 6 females and 14 males. \rv{To be noted, the participants (N=8) who had used KuaiDiGuis independently all have residual visions.} After having the interviews, every participants received 94 CNY as the compensation.

\subsection{Procedure}

%The researchers designed various interview questions to fully investigate the practices and challenges with fetching deliveries of BLV population. 
The interviews followed a semi-structure, and the interviewer asked follow-up questions based on the participants' answers to the interview questions in the following four parts. 
The first part of the interview questions was about the basic demographic information: age, gender, occupation, visually impairment level, assist technology used to operate the electronic devices. 
The second part was about the practices challenges of fetching delivery packages. We asked questions including how often they fetch packages, what the common size and weight of the packages are, how they know that a package arrives, where they fetch the packages, how the participants find the pickup location, what the steps involved in fetching the package are, how the participants identify the belonging of packages, and so on. After asking each practice related question, the researcher asked participants about what specific challenges the participants encounter and how they usually address the challenges. 
The third part was about their attitudes of fetching packages, including what impacts the courier industry have on their daily lives, how they feel about getting help from others, what the concerns are for seeking help from others, what the advantages and disadvantages are of various delivery methods (e.g., door-to-door delivery, collective pickup site, KuaiDigui, package pile), what the participants care about the most in the process of fetching delivery packages (e.g., Independence, security, privacy, efficiency), and so on. 
The final part is about the expectations of fetching packages, including questions like what the ideal situation for the fetching packages process is, and how technology could improve the accessibility level of package fetching process.

\subsection{Analysis Method}
All interviews were conducted face to face by the first author in Mandarin as all our participants were Mandarin native speakers. The researchers used an iOS mobile phone to audio-record all interviews and transcribed the records verbatim into Chinese via Iflyrec (i.e., a voice transcription platform). After having the transcriptions translated in English, two researchers followed an open coding approach~\cite{corbin1990grounded}. They first independently coded the data and then later discussed the codes  to reach a consensus on their interpretation of the data and the codes. They then applied affinity diagramming on the codes using an online tool, Miro, to group the codes into clusters and derive common themes emerging from the data. During the process, two researchers consulted both the English translation and the Chinese transcripts to ensure the accurate interpretation of the data. % \cite{hartson2012ux}.

\section{Findings}
We present our findings on the following key themes: \textit{how BVI users fetch packages},  \textit{how they interact with KuaiDiGuis}, the self-service package pickup machine, and  \textit{their identification as a BLV person while fetching packages}. In addition, we will explain \textit{how BLV people seek help from other people} within the above themes.  

\subsection{How Do BLV Users Fetch Packages?}
For general package-fetching, there were several key processes noted by participants. These included \textit{finding and going to the pickup sites}, \textit{package localization}, and \textit{recognizing and identifying packages}.

% \subsubsection{Logistics Information}
% Participants reported that they mainly query logistics via the online shopping APP, the logistics company's WeChat official account, or searching tracking number on the CEP company's website. 
% These behaviors could be all conducted on mobile phones or computers, so they can easily read the detailed and clear logistics information through the screen reader. 
% \begin{quote}
% ``...For logistics, it did a great job on the screen reader accessibility. After I checked in the order within 1~2 hours, there would be logistic information all the way, such as your package is picked up by courier company, which operator picked it up and so on...'' - P1
% \end{quote}

% Though BLV people can read logistic information via screen reader with ease, there are some challenges when querying logistics. Since the screen reader usually reads information sequentially from top to bottom, P8 (M, totally blind) reported that it would cause confusion when matching logistics status with time information. \textit{"Sometimes the layout of logistics information is a bit confusing when using the screen reader. When it comes to matching logistics information and time information, sometimes I might be confused of the mapping relationship."}, said P8 (M, totally blind).

\subsubsection{Find package pickup sites}
Although packages can be delivered straight to homes or through face-to-face contact, it is more common for packages to require pickup in China, causing people to leave their homes and travel to a pickup site. Some examples include self-service package pickup machine pickup, package station pickup, community office pickup, and convenience store pickup. For BLV people, it is challenging to complete this task independently. \rv{Most}\st{Some} participants reported asking other sighted people such as family members, friends or neighbors to accompany them to the pickup site \rv{(90$\%$, N=18)}. After memorizing the route, participants located these sites with the help of environmental sounds and other unique signals on the route \rv{(45$\%$, N=9)}.
\begin{quote}
P2 \st{(F, totally blind)}: ``...If I want to find the pickup site, I need go to the other building. To tell if I arrived, I can listen to the sound of the escalator next to the target building. If the escalator is working, it will make a sound, that’s when I know I am in distance of the destination. If I want to go to the package station, there is a mahjong club there. The sound of mahjong club is pretty loud and I can tell if I am near there...''
\end{quote}

However, relying on memory is not always effective. \rv{A few}\st{Some} participants reported they might forget how to get to the package pickup site after not taking the route for a period of time \rv{(20$\%$, N=4)} . P3 \st{(M, totally blind)} stated \textit{"I might be familiar with the route at the beginning, but after a period of time, the memory may become a little fuzzy"}. Some participants also mentioned asking strangers for directions to the pickup site when they fail to find them on their own \rv{(30$\%$, N=6)}. P2 \st{(F, totally blind)} mentioned starting a mobile video call with her family members and aiming the camera on her path. This helped family members guide the participant by verbally directing which way to go and when to stop. P2 \st{(F, totally blind)} reported that though the video call method allowed her to get instant help, it caused her to pay less attention to the surrounding environment. 

\begin{quote}
P2 \st{(F, totally blind)}: ``...When I am holding the phone and having a video call, I do not pay attention to any environmental changes around me, and focus only on the call. Also, I would hold the phone a little higher for the helper to gain a broader view and me tell where I am and whether I reached the destination. But I may not be able to recognize changes in elevation, and the helper couldn't see an elevation and alert me about that...'' 
\end{quote}

P5 \st{(M, blind with residual vision)} reported that he also failed to fetch packages because he took the wrong route and went to the wrong pickup site.

\begin{figure*}[htp]
  \centering
  \includegraphics[width=\textwidth]{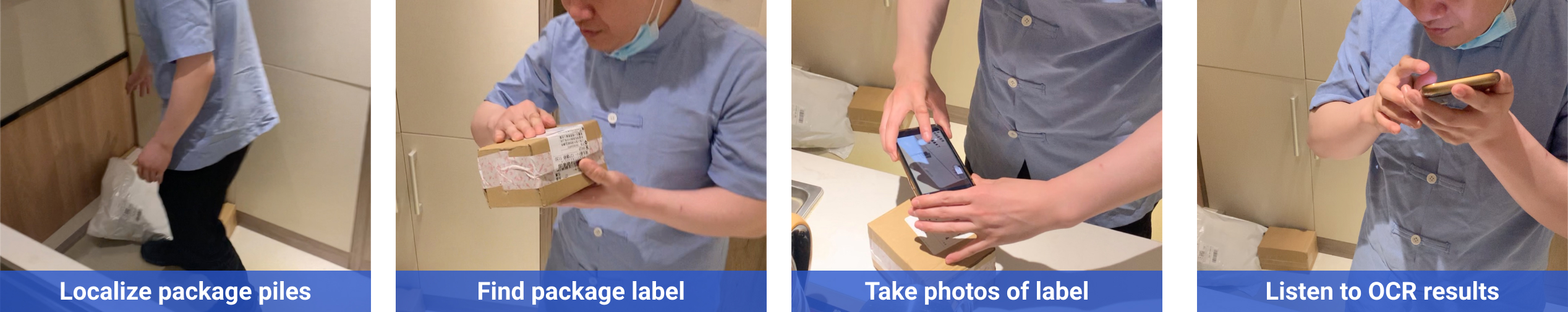}
  \caption{Key steps to localize and recognize a package.}
  \label{fig:L_R_process}
  \Description{Four images that show the process of how a blind participant localize and recognize his package from a packages pile, including package piles localization, package label localization, taking photo of the label, and listen to OCR results from the mobile phone.}
\end{figure*}

\subsubsection{Package Localization}
Participants described some common scenarios for package localization such as the front desk, outside the house door, in the building's corridor, in package piles, and on package shelves.

Low vision participants and blind participants with residual vision reported that if there was a package of a certain size outside their doors, they could localize the package with residual vision \rv{(N=12)}. \textit{"If there was something outside the door, like a carton or a bag, I can recognize it with my residual vision,"} said P5 \st{(M, blind with residual vision)}. Figure \ref{fig:L_R_process} shows how one participant localizes and recognizes a package. This workflow was a common approach adopted by other participants as well. As seen in the first image, \st{the participant} \rv{P1} chose to explore the package location with their limbs or white canes. P12 \st{(M, blind with residual vision)} said \textit{"If it's very close to the shoe cabinet outside the door, I basically think it's okay to use hands. But if it's in a faraway area, I don't dare to use my hands directly. I might take a white cane and explore it"}. \rv{A few} participants reported that they tended to be slow and cautious when localizing packages for safety reasons \rv{(20$\%$, N=4)}. \textit{"I definitely have some concerns when localizing packages by touch. When we touch, our movements are very slow and light. We can only explore little by a little,"} elaborated by P1 \st{(M, totally blind)}. 

% Some participants reported accidentally bumping into unexpected packages and having their doorway blocked.

% \begin{quote}
% ``...If relatively large packages are placed outside the house door, it will affect my movement. I need to move them away so that the door can be opened. Also, when I come back with a bag in my hand, I may choose to put down the things I carry with first, and then I can have hands to move packages...'' - P2 (F, totally blind)
% \end{quote}

The package localization process is also affected by the lighting condition of the environment. \textit{"For me, I still mainly use my vision to localize the packages. But at night, I might have to use a flashlight for exploration,"} said P12 \st{(M, blind with residual vision)}. \textit{"To find a package at the door, I need to turn on the flashlight, because my [residual] vision is better in bright places, it is not easy to perceive things if the lighting is bad, especially if the packages are delivered at night,"} said P14 \st{(M, blind with residual vision)}.

We found that localizing small packages is extremely difficult for most participants \rv{(95$\%$, N=19)}, as they often miss small packages when exploring by the above approach of touching and kicking or utilizing residual vision. \textit{"If it is a big box or a carton, it will be easier to localize. But if it is a small bag packaged with thin materials, I may not notice it too much, and I may think it is garbage bag or something,"} said P14 \st{(M, blind with residual vision)}. Participants reported that they needed to explore the environment more, and would usually turn to sighted people to help find small packages \rv{(85$\%$, N=17)}. To avoid this situation, some participants would take precautions such as confirming the package location with the delivery worker in advance \rv{(25$\%$, N=5)}. 
\begin{quote}
P10 \st{(F, blind with residual vision)}: ``...I will accurately confirm the location with the delivery worker and tell him that our house is inside an iron door, and he can place the packages on the window on right side, or place it next to or under the tile on the ground...'' 
\end{quote}

\subsubsection{Package Recognition}
After localizing the package, the next step would be package recognition, as shown in Figure~\ref{fig:L_R_process}. This requires BLV people to know the owner of the packages. The recipient information is often stored on the delivery label of packages, which is hard for BLV people to read as quickly as sighted people. In the interview, we found that BLV participants have their own ways to perform package recognition and also encountered challenges during the process.

When recognizing packages, some participants mentioned that they usually know the content of the package, and make a preliminary judgment before opening it based on the size, material, and weight of the packaging \rv{(45$\%$, N=9)}. 
\begin{quote}
P13 \st{(M, low vision)}: ``...You need to know what to pick up. For example, I’m going to pick up a large bottle of laundry detergent. First of all, I know that the item is very heavy. It may be packed in a box. After getting those larger boxes, I will shake it. If there is no liquid sound inside, [for example] instant noodles which are quite heavy and packed in boxes which rustle with the sound of dry food when shaking it, then I can be determine that it is not my laundry detergent...''
\end{quote}

Compared to recognizing a single package, it would be more challenging when recognizing a single package among a group of packages via size, sound and weight. \textit{"[If]there are a lot of packages piled up and they are about the same size., you need to find someone sighted to help you look for it,"} said P4 \st{(F, blind with residual vision)}.

An important step to recognize whether the package belongs to them was to recognize the delivery information label outside the package. The first step was to identify the location of the label. As seen in the second image of Figure \ref{fig:L_R_process}, for participants who are totally blind \rv{(N=8)}, the main identification method reported was through feeling texture differences via tactile feedback. Participants with residual vision \rv{(N=12)} also reported that the color difference between the packaging (usually Brown) and the label (usually White) was also an important reference for judgment. 
\begin{quote}
P13 \st{(M, low vision)}: ``...Most packages will have a text area with a white background which is the delivery label. In most cases, the recipient, cargo, and logistics information are in that text area. First, the feel of the cardboard box is relatively rough and the feel of the delivery label is a bit smoother. Second, it can also be identified by my residual vision because the color difference is obvious between boxes and delivery label. The carton is usually brown, and the label is usually white and has black letters on it...'' 
\end{quote}

After confirming the location of delivery label, the next step was to read the information on it. During the interview, participants highlighted two main methods to accomplish this independently. Participants \rv{(N=12)} with residual vision and text recognition used smart phones to zoom in on the image taken from their camera. Then they read the information on the packaging directly by looking close to the enlarged images. As seen in the third image of Figure \ref{fig:L_R_process}, totally blind participants \rv{(N=8)} would take a photo of the delivery label with their mobile phones, use the Optical Character Recognition (OCR) feature to recognize text from the photo, then read the information via screen readers. However, the reading label process was full of challenges. Participants reported needing to be precise when aiming the camera at the delivery label when using OCR, if the camera is not framed correctly on the label or if the photo taken is blurry, OCR may process incorrectly \rv{(30$\%$, N=6)}. 
\begin{quote}
P1 \st{(M, totally blind)}: ``...I have to touch the label with one hand, pick up the phone with the another hand, keep a certain distance from the label, align the direction and then take the picture. Sometimes it may be inaccurate like take pictures of the wrong parts. If one shot is unsuccessful, I will have to take multiple shots...''
\end{quote}

\begin{figure}[ht]
  \centering
  \includegraphics[width=\linewidth]{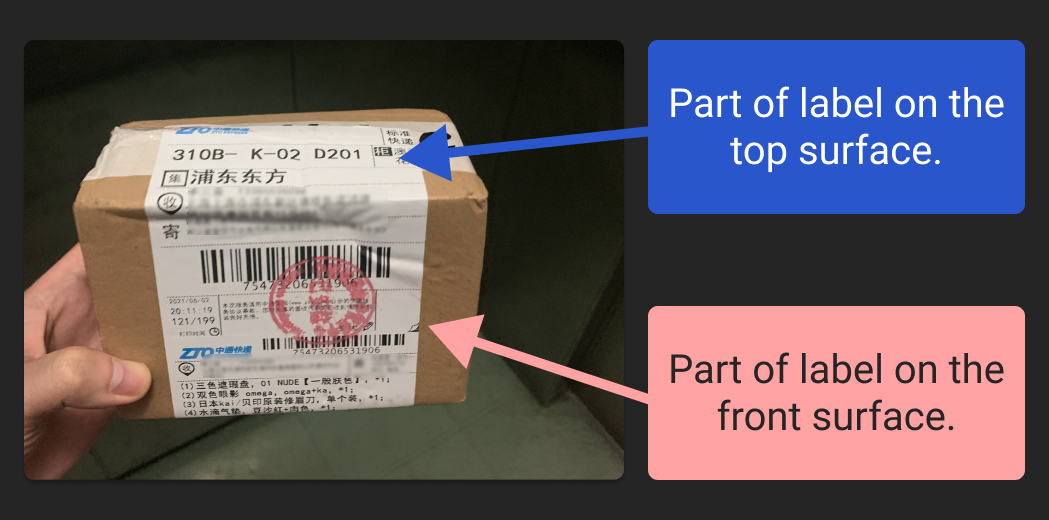}
  \caption{Label on different surfaces of the package.}
  \label{fig:multipleSurface}
  \Description{An image that shows a package wrapped by a large delivery label which have part of the label on the top surface of packaging and part of it on the front.}
\end{figure}

As seen in Figure \ref{fig:multipleSurface}, a small package with a large delivery label causes the label to span multiple sides of the package. Some participants reported not being able to take a single photo of the whole label which may contain valuable information (e.g., name, address) \rv{(45$\%$, N=9)}. Irrelevant information such as advertisements and slogans would be captured instead, which would be recognized by the OCR feature and causes the user to take multiple pictures. 

\begin{quote}
P5 \st{(M, blind with residual vision)}: ``...Some cartons have a lot of text on them, and I have photographed some information that has nothing to do with the recipient. They are just some product introduction. In this case, I will find other similar label areas and take another shot...''
\end{quote}

Furthermore, P6 \st{(F, blind with residual vision)} complained that using a screen reader to read the recognized information was inefficient because of the fixed reading sequence. \textit{"The system would read something captured from the photo that I don’t want to know. Also, it reads from beginning to end, and I can’t intercept part of it. If I didn't understand or miss part of the content, I have to listen again from the first sentence,"} said P6 \st{(F, blind with residual vision)}.

Participants with residual vision mentioned that they would take photos of labels and zoom in to see recipient information. However, participants had low Chinese literacy rates and most could only recognize their names. Other texts (e.g., sender’s name, address, courier company name) on the delivery label impeded the result in finding their names. \textit{"...The characters are still too small, they are blurred when zoomed in, and I don't know many Chinese characters other than my name. And it’s hard to find my name on the delivery label, because there are some other information on it,"} said P5 \st{(M, blind with residual vision)}.

Many participants mentioned that the picture recognition process was too complicated because it took a lot of time and often required multiple attempts \rv{(65$\%$, N=13)}. This made users not want to perform OCR on the packages. 

\begin{quote}
P6 \st{(F, blind with residual vision)}: ``...It takes a long time. The recognition function still has many flaws. Once [I] turned the recognition on, other features would become unavailable. And you have to turn [the recognition function] off to use the mobile phone normally, which is troublesome...''
\end{quote}

\subsubsection{Seeking Help}
All participants reported that they would turn to people with better vision for help when having trouble localizing or recognizing packages alone. \textit{"Not all of them are sighted people, they could be someone who has better vision than me, such as some of my low vision friends,"} said P13 \st{(M, low vision)}. Participants \rv{(80$\%$, N=16)} tended to choose those whom they are familiar with, such as family members, colleagues, neighbors, friends, etc, because they believed that these people could better understand their situations. \textit{"Because my colleagues know me better, and for people who don’t know me, they may not understand that I am visually impaired and have more questions to ask about me, which will cost extra time,"} explained P6 \st{(F, blind with residual vision)}.

However, there are multiple challenges when seeking help. Helpers like colleagues and family members might not be available all the time. \textit{"My colleagues are actually very busy and have their own jobs, and I don’t want to trouble them,"} said P3 \st{(M, totally blind)}. When fetching packages alone, some participants reported that they might not be able to find other people present to ask for help \rv{(35$\%$, N=7)}. \textit{"I really couldn't do anything to find the package on my own, so I had to wait for almost half an hour until someone came over to ask him to find it for me. This can only depend on luck,"} said P5 \st{(M, blind with residual vision)}. A few participants mentioned they would seek online help (e.g., having video calls with family or friends via WeChat, using a real-time helper APP to connect with sighted helpers) \rv{(20$\%$, N=4)}. \textit{"I asked my family to help me look for packages through the WeChat video call. I aimed the camera to this label and ask them to read the details for me,"} said P7 \st{(M, blind with residual vision)}, who also mentioned the online help approach required a good video or photo quality for the sighted helper to identify the information. 
\begin{quote}
P7 \st{(M, blind with residual vision)}: ``...I video called my wife first, but she said that the video was not clear, and she couldn't confirm which was mine after going through several packages Finally, I asked a passerby to help me check it out...''
\end{quote} 

P7 \st{(M, blind with residual vision)} also mentioned that he did not identify the packages right after the sighted passerby helped him; instead, he waited to do so in private. \textit{"I took the package to a place with no one around and recognize it. I was embarrassed to confirm it in front of the helper. One reason is that since I just asked him to help, and then if I still confirm it again, it will make the helper feel that he is not helpful or I don't trust him,"} said P7 \st{(M, blind with residual vision)}.

\subsection{Usage of KuaiDiGui}
KuaiDiGUis, a form of self-service package pickup machines,  have been mushrooming in recent years in China. Many packages are no longer delivered to the door as delivery workers send packages to KuaiDiGuis by default. To pickup packages from KuaiDiGuis, there are six main steps: pickup preparation, arriving at a KuaiDiGui, operating the KuaiDiGui’s touch screen, finding the opened compartment door, taking out the package, and then closing the door. In the interview study, participants reported challenges they encountered and their solutions when using KuaiDiGuis.

\subsubsection{Pickup preparation}
Participants mentioned that they would bring mobile phones and white canes with them when picking up packages from KuaiDiGuis. Mobile phones are necessary since BLV people need to get the pickup code or use the camera to scan a QR code. A few participants mentioned that they would schedule extra time for KuaiDiGui pickup because it takes longer timer to operate the machine \rv{(10$\%$, N=2)}. \textit{"You may need to reserve more time for KuaiDiGuis, because you have to find the location first and explore how to use it. Some errors with the system might occur, like the door did not open or no feedback when scanning the QR code,"} said P5 \st{(M, blind with residual vision)}.

\subsubsection{Distinguish different KuaiDiGuis}
\begin{figure}[htb!]
  \centering
  \includegraphics[width=\linewidth]{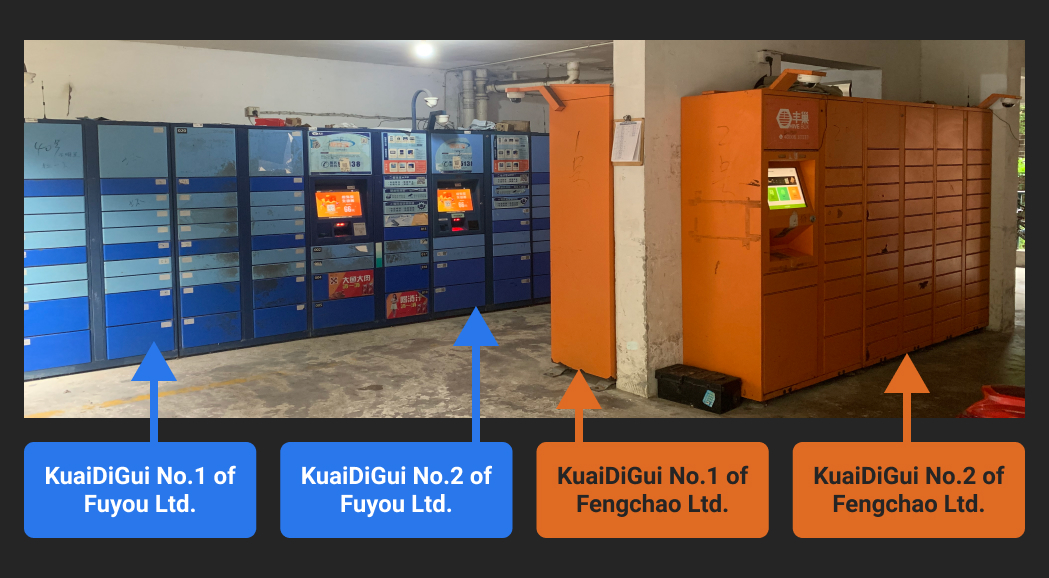}
  \caption{Four KuaiDiGuis from two different companies in the same area.}
  \label{fig:multipleKDGs}
  \Description{An image that shows four KuaiDiGuis from two different companies in the same area.}
\end{figure}
As seen in Figure \ref{fig:multipleKDGs}, after the BLV people arrive at a KuaiDiGui site, they need to identify the right one because there could be multiple KuaiDiGuis with different numbers and different brands in the same area. Some participants said that they would choose to try KuaiDiGuis one by one by entering pickup code to see if any compartment’s door would open \rv{(35$\%$, N=7)}. This method is limited to participants who have residual vision and can independently operate KuaiDiGuis \rv{(N=8)}. For totally blind participants \rv{(N=8)}, they reported that they needed other people's help to distinguish different KuaiDiGuis. \textit{"I have to ask passerby for help. I have no way to distinguish [KuaiDiGuis] by myself,"} said P11 who is totally blind\st{(F, totally blind)}.

\subsubsection{Identify elements on the touch screen}
Identifying elements like buttons, keyboards, text hints etc. on touch screens is an essential part of manipulating the KuaiDiGui system. However, most participants reported that due to the lack of screen reader feature and voice feedback in the KuaiDiGui system, they had little experience on operating KuaiDiGui or failed to operate it independently \rv{(75$\%$, N=15)}. \textit{"There is no voice prompt for KuaiDiGui. I cannot operate it at all,"} said P12 \st{(M, blind with residual vision)}. On the other hand, all participants who had experiences using KuaiDiGuis alone had residual vision, and could recognize images and texts by getting closer to the interface or by enlarging it via camera zoom \rv{(40$\%$, N=8)}. The description of how to use KuaiDiGuis' touchscreens independently in this study were all from the participants with residual vision. 

% \begin{figure}[h]
%   \centering
%   \includegraphics[width=\linewidth]{supplements/KDG pickup method.jpg}
%   \caption{Scan QR code or enter pick-up code to pick up.}
%   \label{fig:codeinterface}
% \end{figure}

As seen in the third image of Figure \ref{fig:FCinterface}, there are two main ways to accomplish pickup process for KuaiDiGui. One is to scan a QR code, the other is to enter a 6 or 8-digits code on the touch screen keyboard. Both methods required BLV people to go through the task flow on the touch screen by clicking the right buttons (e.g., “Take Express" button, “Pickup by pickup code” button, numeric key buttons). Identifying the position and function of the elements is inevitable. Low vision participants and participants with residual vision reported either getting as close as possible to the screen or taking a picture to zoom in on, then going through the patterns and text little by little \rv{40$\%$, N=8)}. \textit{"If the words on the screen of KuaiDiGui are relatively large, I can see them by getting closer. If I can’t, I will take a picture and zoom in,"} said P10 \st{(F, blind with residual vision)}.

Most participants mentioned knowing little written Chinese characters \rv{(75$\%$, N=15)}. \textit{"I mainly learn Braille and don’t know much about Chinese characters. For KuaiDiGui, I can see there’re some patterns on the screen, such as the pattern of a box for "pick up" function,"} said P5 \st{(M, blind with residual vision)}. Some participants used colors to tell different buttons \rv{30$\%$, N=6)}. \textit{"Although the characters are small, I can see that this area is yellow, and that area is green. I will remember the functions corresponding to each color area,"} said P9 \st{(M, low vision)}. However, even with these approaches, the process was still full of challenges. P4 \st{(F, blind with residual vision)} reported that he could only operate the system via the memory of the buttons locations and complete the process by repeated trial and error. P16 \st{(M, blind with residual vision)} reported that he tried to click on the screen, but the outdoor light was very strong, and he failed to see anything because of the reflection on the glass. 

\subsubsection{Error avoidance and correction when entering pickup code}
Participants also pointed out the lack of voice feedback to provide information about if they made an error and how to fix it \rv{(80$\%$, N=16)}. \textit{"If the input is wrong, and the KuaiDiGui also does not broadcast it, I would not know that,"} said P5 \st{(M, blind with residual vision)}. When entering the pickup code, a few participants reported that even if they realized they entered a wrong code, they would not try to find the Delete button, which was challenging, but instead continue to enter random digits until the system announced the code was wrong and reset the process for them to start over \rv{(20$\%$, N=4)}. %even they knew the code they entered was wrong to avoid finding the delete button. 
\begin{quote}
P9 \st{(M, low vision)}: ``...To delete the input, I have to find the delete button location, which may be more troublesome. I will enter all numbers in one time even if I know I click on the wrong one, because the system will let me re-enter pickup code again if the first round input is wrong. In this way, I don’t need to find the delete button on purpose...'' 
\end{quote}

\begin{figure}[ht]
  \centering
  \includegraphics[width=\linewidth]{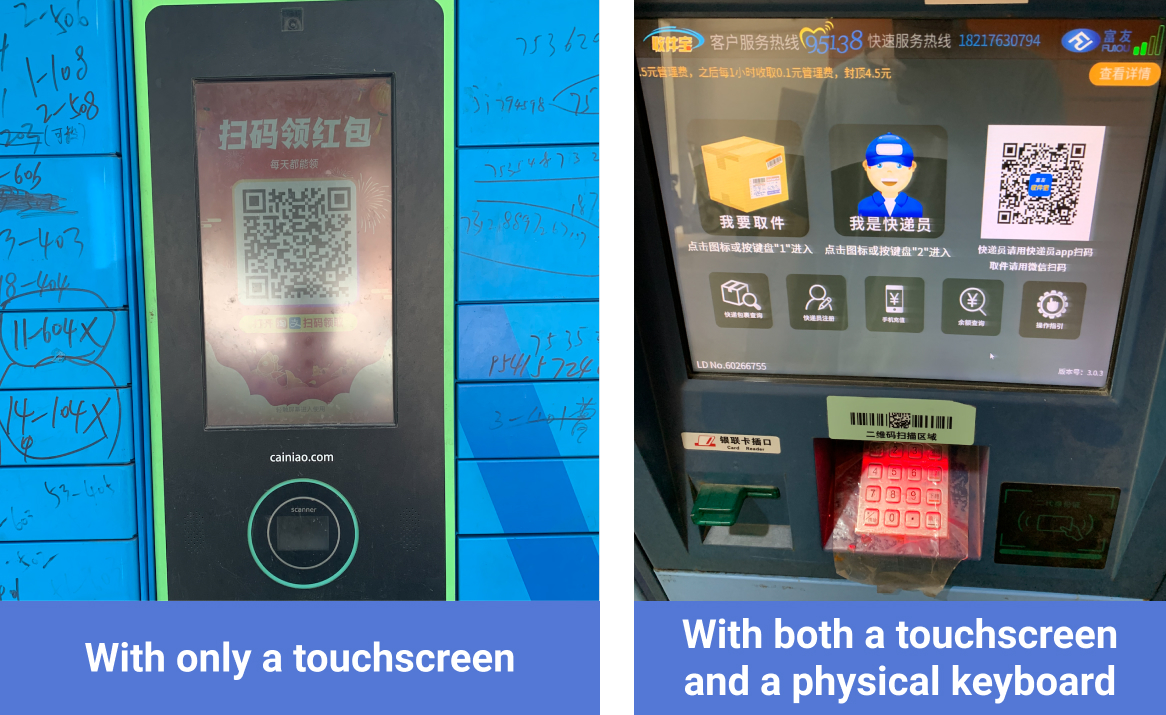}
  \caption{Touch screen and physical keyboard on KuaiDiGuis.}
  \label{fig:KDGkeyboard}
  \Description{Two images that show different interactions on different KuaiDiGuis, including KuaiDigui with only a touchscreen, and a KuaiDiGui with both a touchscreen and a physical keyboard.}
\end{figure}

As seen in Figure \ref{fig:KDGkeyboard}, some KuaiDiGuis were equipped with both a touch screen and a physical keyboard. A few participants also reported that they would like to use physical keyboards to enter pickup code on KuaiDiGuis(15$\%$, N=3). \textit{"There will be a bump on the number 5 button in the middle of the physical keyboard, so that we can know what key it is when touching it. Then I can enter a pick up code on the physical keyboards by memory of the key distribution,"} said P10 \st{(F, blind with residual vision)}. However, many participants said that they did not realize that there was a physical keyboard because the smooth large touch screen was the most conspicuous element on the machine(50$\%$, N=10). 

\subsubsection{Complex KuaiDiGui system}

Low vision participants \rv{(N=2)} preferred a KuaiDiGui system with a short task flow and concise interfaces. \textit{"The steps on the touch screen should be as few as possible. Take the Fengchao KuaiDiuGui as an example. You first need to find the ‘pickup’ button and click it, and then scan the QR code. I think it’s a bit complicated, which is difficult for totally blind users,"} said P9 \st{(M, low vision)}. KuaiDiGui systems from different companies have different interfaces, which caused more troubles to a few participants \rv{(15$\%$, N=3)}. \textit{"I'm not sure about the element changes on the touch screen. The position of the keyboard is not fixed. Sometimes it appears at the bottom of the screen, and sometimes it appears in the center. Also, the content of the keyboard is also not fixed, sometimes it has 26 English letters, but sometimes it has 9-digit number,"} said P3 \st{(M, totally blind)}. For the pickup code consisting of both numbers and letters, P4 \st{(F, blind with residual vision)} said she would not fetch the package by herself, because she felt the keyboard was too complicated for her to operate. Participants also mentioned the unexpected pop-up advertisement on the KuaiDiGuis' touchscreens\rv{(20$\%$, N=4)}. 
\begin{quote}
P10 \st{(F, blind with residual vision)}: ``...The Fengchao KuaiDiGui will suddenly pop up an advertisement. I don’t know where to turn it off because the cross or skip buttons are too small to find. I might be stunned there for a while and don't know what to do, and have to wait until someone comes and ask them for help...''
\end{quote} 
P14 \st{(M, blind with residual vision)} mentioned that he would wait for a while until the advertisement disappeared. 

\begin{figure}[ht]
  \centering
  \includegraphics[width=\linewidth]{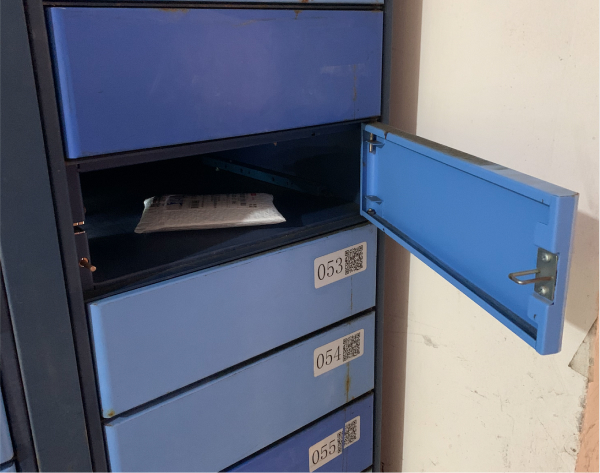}
  \caption{The opened door of the compartment}
  \label{fig:compartmentdoor}
  \Description{An image that shows a package in an opened compartment of KuaiDiGui.}
\end{figure}

\subsubsection {Identify the opened compartment on KuaiDiGui}
After the touch screen, the next step was to find the opened compartment seen in Figure \ref{fig:compartmentdoor}. A crisp sound is heard when the door opens, and participants reported that they could tell the general direction of the sound, but still need some guidance to get to the exact opened door as there were many compartments in any given direction \rv{(60$\%$, N=12)}. \textit{"I will first listen to the sound to determine the direction, I will hold on the surface of KuaiDiGui, and then touch the opened door with my hand,"} said P15 \st{(M, blind with residual vision)}.

Finding the opened compartment with sound does not work all the time. P5 \st{(M, blind with residual vision)} reported that if the sound was quiet or if the door only opened a little, it was difficult to determine the location. Some participants mentioned that in the process of feeling around, they would be very cautious and avoid fumbling and kicking because they were worried about accidentally closing the door \rv{(35$\%$, N=7)}. \textit{"I won’t kick around to find the door, because it will be troublesome if I accidentally kick the door and close it,"} said P5 \st{(M, blind with residual vision)}.

Participants also mentioned the dangers of an opened compartment door \rv{(30$\%$, N=6)}. One issue was that BLV people might collide with the door. P4 \st{(F, blind with residual vision)} and P5 \st{(M, blind with residual vision)} both reported that they had collided with the door opening beside their heads during the exploration process. The materials of KuaiDiGuis also raised another safety issue. \textit{"There would be rust on KuaiDiGuis if they were outside for a long time. When touching it, I worry that if there is something sharp is rusty, it might cause tetanus if it cuts my hand,"} said P15 \st{(M, blind with residual vision)}.

\subsubsection{Seeking help around KuaiDiGuis}
Due to the challenges of using KuaiDiGuis, many participants reported that they would turn to sighted people (e.g., neighbors, passers-by, or other strangers) for help when dealing the KuaiDiGui alone \rv{(85$\%$, N=17)}. However, the assistance from sighted people often relies on timing. Many participants mentioned that if the timing was not right and no one else was around, it would take a while to get help \rv{(70$\%$, N=14)}.
\begin{quote}
P3 \st{(M, totally blind)}: ``...I will try to [catch] people, and ask them to help pick up for me. I usually wait there around dinner time and there would be lots of people. If it is two or three o'clock in the afternoon, there may be no people to [catch] ...''
\end{quote} 

In terms of helper selection, some participants preferred those who were younger, had certain education, and were familiar with the use of electronic products \rv{(25$\%$, N=5)}. \textit{"I usually choose the younger ones because they were easier to communicate. On the other hand, people in their 50s and 60s may not be familiar with the operation of KuaiDiGuis,"} said P20 \st{(M, totally blind)}. 

When asking helpers to accomplish the full picking up process, it is necessary for the BLV people to provide the pickup code. Participants reported two common methods. One was to verbalize the pickup code to the helper which required BLV people to remember the pickup code in advance or listen the code via the phone on site. Another way was to directly show their mobile phone with the pickup code page to the helpers. Some participants said they might hand over their mobile phone to the helpers. \textit{"I will display the pickup code on the phone, and turn off the screen reader mode, then give it to the helper. In this way, it is easier for him to operate my phone,"} said P2 \st{(F, totally blind)}. Most of the participants stated that they had no safety concerns when displaying their mobile phones but would not let helpers control the phones \rv{(50$\%$, N=10)}, \textit{"I don't have any worries, because I only display the pickup code page to him, and I won't let him operate the phone,"} said P3 \st{(M, totally blind)}.

\subsection{Attitudes toward unmanned technology}
\rv{KuaiDiGuis are the most mature representation of an unmanned delivery product. Other types of unmanned technologies, such as delivery vehicles and drones, are emerging on the market. We asked participants about their attitudes and expectations towards such technologies. Their general impressions were that accessibility issues in the package delivery process were often not considered during the development of unmanned technology.} 

\rv{Though some participants believed that unmanned delivery vehicles could improve the efficiency of delivery (30$\%$, N=6), many participants have concerns on whether they can operate them independently due to their previous inaccessible KuaiDiGui pickup experience (65$\%$, N=13).}\textit{\rv{"Personally, I am looking forward to the application of futuristic tech for delivery, but it should definitely include accessible designs. Otherwise, the advances in technology would only bring more obstacles to us, just like the KuaiDuiGuis}} \rv{said P2} \st{(M, totally blind)}. \rv{Some participants worried about the difficulty in seeking help while interacting with the unmanned delivery machines (35$\%$, N=7).} \textit{\rv{"If it is an unmanned pickup machine, I can only wait for help passively when I can handle some problems myself. But if there is someone on-site, I can definitely communicate with them and seek help."}} \rv{said P8} \st{(M, totally blind)}.

\rv{Although participants have some accessibility concerns for unmanned deliveries, this does not mean that the BLV community completely rejects new technologies. Participants still expressed strong expectations toward future unmanned delivery methods.} \textit{\rv{"It would be super cool to have drones or auto-driving cars to send my packages directly to my door or window, which can save my time and effort to go to the collective pickup sites,"}} \rv{said P12} \st{(M, blind with residual vision)}. \rv{Many participants expressed their willingness to embrace unmanned express delivery technology under the premise of accessibility functions. P13} \st{(M, low vision)} \rv{reported that if the KuaiDiGui can be used like a mobile phone via screen readers, he would definitely use it as it solves the problem of searching and recognizing a large number of packages}. 

\subsection{Self-identity reflection and thinking}
Some participants admitted that because of the visual disability, they were unable to do something totally independently \rv{(30$\%$, N=6)}. \textit{"As a blind person, it would be difficult to be completely independent without turning to someone for help,"} said P1 \st{(M, totally blind)}. On the other hand, most participants said they did not want to trouble others too much, and they would first try to do something by themselves before asking for help \rv{(90$\%$, N=18)}. This could help them develop the feeling of independence, the sense of accomplishment, and increase connection with the outside world. 
\begin{quote}
P3 \st{(M, totally blind)}: ``...Some visually impaired people are afraid to leave their house and communicate with people outside, I think it is good to walk around to pick up the packages in the community, which not only allows BLV people to exercise their bodies, but also improve the ability to live independently...''
\end{quote} 
P12 \st{(M, blind with residual vision)} believed that the independence was most important for fetching packages. \textit{"My wife and I are both completely blind. Distant water won't quench present thirst, the help from others is not available all the time, so we need to live independently,"} said P12\st{(M, blind with residual vision)}. A few participants also had concerns about seeking independence \rv{(20$\%$, N=4)}. When trying to use KuaiDiGuis or other pickup methods independently, P5 \st{(M, blind with residual vision)} and P10 (F, blind with residual vision) felt that she often took far more time than sighted people, and she worried about whether her prolonged operations would affect others' pickup. 

Most participants mentioned that they did not mind others knowing they were visually impaired when fetching packages \rv{(85$\%$, N=17)}; they would even explain their vision ability to others to get help. However, for P13 \st{(M, low vision)}, he reported his concerns about his BLV identity. 

\begin{quote}
P13 \st{(M, low vision)}: ``...I don’t like to use obvious assisting tools like electronic vision aids, screen reader, and white cane outside my home or the school when fetching the packages, because I am worried the neighbors in the community would look at me and my family differently. Also, if I turn to someone unknown for help, I need to explain my visual impairment to him, which is embarrassing...''
\end{quote} 
Other participants also mentioned that when communicating with the courier companies, they tended to not indicate they are visually impaired and did not want to be treated differently than sighted customers \rv{(40$\%$, N=8)}. They would only confess their visual impairment if a delivery worker asked them to accomplish a vision related action.

\section{Discussion}
In the interview study, we found that major accessibility issues occurred during the package localization and recognition process while interacting with KuaiDiGuis. We discuss the issues during these processes, how previous research could address them, and provide areas for improvement for BLV people's package fetching process.

\subsection{How to make the package localization and recognition process more accessible?}
Package localization and recognition are key steps for package fetching. We discuss the main pain points during the process, how assistive technology proposed in prior studies could help BLV users with indoor navigation, and how object identification might be able to address these pain points and remaining challenges.

\subsubsection{Package localization}
We found that to localize packages, participants needed to rely on their limbs, residual vision, or white cane to explore their environment. \rv{Prior studies on} indoor navigation assistive technologies for BLV people mostly focused on helping them navigate in an indoor environment by showing additional information on a map \cite{gleason2016vizmap} or projecting light indicators on stairs (e.g., \cite{,yuhang2019AR}). However, our study shows that localizing packages requires BLV people to perform a fine-grained exploration of a small area, which can be particularly challenging if the target package is small, crowded with similar sized packages, or when the area has safety concerns.  
%Previous studies investigated how to detect object characteristics, position, orientation and navigation with sensors \cite{Andreas2003Indoor}. 
\rv{During such an exploration, contextual information about the objects (e.g., the size, shape, location, and number of objects) and the area (e.g., obstacles) can be instrumental to BLV people. Utilizing a device with computer vision that provides auditory or haptic feedback can be a method to localize packages and guide BLV people to their package.}
%Cole et al.'s VizMap \cite{gleason2016vizmap} might be a useful approach allowing for labeling additional information on items. 
However, it remains largely an open challenge how to help BLV people safely and effectively locate a target package, in particular to discern a small package crowded in similar sized packages.

% %, designed the VizMap system to help visually impaired people explore visual information in indoor environments by labeling the items information in a reconstructed 3D spatial model. \cite{gleason2016vizmap}. Cole’s approach could also be applied in the package finding process by labeling the packages and supporting BLV people perceive information via smartphone or white cane as they move in the environment.

% AR technology can also be applied during the package localization process for low vision people. Yuhang et al. proposed an AR visualization technique that projects colored labels on stairs through head-mounted AR to help low vision people navigate stairs in their environment. Image processing technology can help to enhance the outline of packages Those with low vision can easily find packages in the environment, especially packages of a smaller size \cite{yuhang2019AR}. However, the AR approach cannot be applied to the totally blind population because it requires the users to have residual vision.

\subsubsection{Package recognition}
To recognize packages, participants relied on tactile feedback (e.g., hand touch) and residual vision to locate the printed label and to recognize information on the label. Participants had to aim their phone's camera at the label to capture such information and either zoom in to recognize the label with their residual vision or use OCR to recognize the label. This proved to be a challenging process for our participants, in particular when the label wraps around two surfaces of the package. They often aimed at the wrong places, partially captured the label, or took blurry images. Consequently, participants reported needing to take a few shots continuously. 
%Prior research investigated methods to extract high-quality photos from continuous camera video for object recognition\cite{yu2013realtime}. While this approach allows BLV people to recognize the object without aiming at the target precisely, it does not help recognize text-based information, which would not be applicable for label recognition. 
Prior research proposed methods to help BLV people recognize objects by automatically extracting a high-quality photo \cite{yu2013realtime,dragan2020ReCog}. While such object recognition approaches might be helpful for BLV users to recognize a package or a printed label area, it remains a challenge to help them aim the camera and capture pictures that are sufficient for OCR to recognize texts on it.

% Recently, Dragan et al. designed a camera-aiming guidance system to help blind users adjust camera angles and use deep network algorithm to identify taken photos\cite{dragan2020ReCog}. This approach could be potentially With the aiming guidance technology, BLV people can target the label more efficiently. 
%Prior research proposed a wireless camera/scanner approach to recognize text information. 
To address the package recognition challenge, we might draw inspiration from prior work. Ohnishi et al. proposed a system consisting of a computer, a wireless camera/scanner, and an earphone for blind people to get character information from the environment \cite{Noboru2013Character}. They tested the system in a store scenario and extracted information such as product name, price, and best-before/use-by dates from the images labels on merchandise \cite{Noboru2013Character}. In terms of delivery label recognition, there are also various type of information on the label. This includes the sender and recipient name, address, and contact number. 
However, another key challenge we discovered was that the surfaces of a package often had other distracting texts, such as advertisements. Thus, how to recognize and filter the irrelevant information during the recognition process so that only useful information is delivered to BLV users remains an open challenge. 

Not all our participants could recognize Chinese characters on labels even if OCR and the screen reader could successfully recognize them and speak back to them. One potential low-tech solution, proposed by  
%Chinese characters could be a challenging issue for the current “scan then recognize” approach. 
P13 \st{(M, low vision)} was to print both Chinese characters and corresponding Braille on the label. In addition, providing a tactile label accessible to touch might be another viable future direction. 
%to reading recipient information rather than by sight.

\subsection{How to make KuaiDiGui more accessible for BLV people?}

\rv{We present potential design solutions to improve the accessibility of KuaiDiGui based on our findings and discuss them in the light of the literature. These solutions include installing a screen reader, supporting multi-sensory feedback, using smart phones for interaction, designing simplified task flows and accessible UI for the interfaces, and using facial recognition for identity verification instead of entering pickup code on the screen. Last but not least, we will also discuss about the limitations for each approach.}

\subsubsection{Install Screen Readers to KuaiDiGuis}
Touchscreens are a major way for people to interact with mobile devices and operating KuaiDiGuis. Installing screen readers to all KuaiDiGuis might seem a straightforward solution. However, the screen reader approach also has its limitations. First, it would greatly increase the cost for the KuaiDiGui company. Many participants are worried that the KuaiDiGui company may not be willing to make changes only for BLV customers. Second, after turning on screen reader mode, the operation gestures would change accordingly, which sighted users are generally unfamiliar with and might cause trouble to them if they accidentally trigger it. 
To address this issue, P15 \st{(M, blind with residual vision)} indicated that the KuaiDiGuis can have a fixed button to control the screen reader mode only for BLV people to know. The system could also automatically turn off the screen reader after a period of no operations with the system. Third, when screen reader mode is on, verbalizing all information and actions might cause privacy and safety issues. KuaiDiGuis are usually set in public areas and anyone nearby would hear the information including the pickup code and the compartment location. P13 \st{(M, low vision)} concerned that using a screen reader in a public setting distinguishes the BLV people from sighted ones, causing those nearby to think of the BLV people differently. Nicolau et al. introduced UbiBraille, a vibrotactile reading device, to help blind people read textual information by generating Braille characters \cite{nicolau2013braille}, which could make the KuaiDigui usage process more private and inconspicuous for the BLV population. Nevertheless, further investigation is needed to address the limitations of using a screen reader on KuaiDiGuis.

\subsubsection{\rv{Support Multi-sensory Feedback} to KuaiDiGuis}

We learned from the study that many participants believed having a physical keyboard with 9-digits buttons on KuaiDiGuis could help BLV people use the machine more easily.

We argue that a physical keyboard might not be a perfect solution. Our study suggested that participants needed to perform operations on the touch screen (e.g., clicking “pickup” button, choosing “enter the pickup code to pickup”) before entering the code via a physical keyboard. The traditional 9-digit numeric physical keyboard can only support numeric input, while other operations still need to be completed on the touch screen. Some participants expressed their concerns about this approach. \textit{"It’s convenient to have the physical keyboard, but I feel it’s going against the technology trend. The population it covers is relatively small, which is not a solution for the future. This approach still separates BLV and the elderly from common users by making a special design instead of including the accessibility feature in the whole system,"} said P9 \st{(M, low vision)}.
One line of research addresses these issues by adding tactile interfaces to the touch screen to make it accessible~\cite{guo2017facade,daniel2018braillerinput,zhang20183Dtactilem}. Trindade et al. proposed Hybrid-Brailler to support physical braille input on the back of mobile devices as well as gesture interaction on the touchscreen. This had a better performance with input speeds and accuracy \cite{daniel2018braillerinput}. Zhang et al. presented Interactiles using 3D-printed hardware interfaces to visually impaired users to manipulate mobile apps \cite{zhang20183Dtactilem}. 
\rv{Furthermore, Bidarra et al. designed an interactive kiosk prototype with multiple accessible features (e.g. screen reader, text magnifier, tactile mouse) and tested it with 13 low vision participants. Results showed having multi-sensory features was a more appropriate and efficient solution than simply having a screen magnifier} \cite{KioskPrototype2014}.
\rv{However, such studies exploring tactile or multi-sensory feedback were either focused on touchscreens or for a particular type of vision impairments (e.g., low vision). It remains an open question whether such multi-sensory feedback is appropriate for both blind and low vision people to explore the KuaiDiGui system and how to design such feedback.}

\subsubsection{Use Smart Phones to Interact with KuaiDiGuis}

With the rapid development of 5G technology, KuaiDiGuis have the ability to connect with mobile phones. If all operations currently needed to be done on KuaiDiGui’s touchscreen could be transferred to a smart phone, the BLV people can avoid using inaccessible touchscreens that have no screen reader feature. Guo et al. presented multiple applications to make inaccessible interfaces accessible without adding screen readers to the original interfaces. Their approach was to frame the original interface with mobile cameras, and use a crowd system to label it and describe interface elements beneath users’ fingers \cite{anhong2015appliancereader, Anhong2016vizlens, anhong2017realworldinterfaces, anhong2019statelens}. With this approach, BLV people can use their mobile phone’s screen reader to read out the elements on KuaiDiGui’s touchscreen instead of installing screen readers for all KuaiDiGuis. The advantage of this approach is that the accessibility of smart phones is well developed and smart phones are used by BLV people every day, giving a lower learning cost and less expensive software and hardware to develop.

\subsubsection{Design Simplified Task Flows and Accessible UI for Interfaces}

A shorter process in operating KuaiDiGuis creates a friendlier and easier user experience for BLV users who can interact with the touch screen. Participants with residual vision reported that they would look extremely close to the KuaiDiGui’s touch screen to identify different elements. Important buttons should be enlarged on the interface and the color contrast with the background should be strengthened. For keyboard input efficiency, Rodrigues et al. explored how the sizes of the keys influence blind people’s text entry performance. Results showed that target sizes greater than 10 mm do not improve the input efficiency, and keys less than 2.5 mm come with additional challenges \cite{rodrigues2016keysize}. This research is limited in that Rodrigues et al. conducted tests on a Samsung Galaxy Tab 2 with different size of letter keys. In KuaiDiGui scenarios, users need to enter a numeric pickup code.

Some participants reported that they did not know how to identify written Chinese characters, making text elements play a limited role in the KuaiDiGui pickup process. Different buttons on the touch screen should be distinguished by specific colors or icons to help BLV users remember functions. Finally, the brightness of the KuaiDiGui’s touch screen should be high enough to avoid being unclear under strong sunlight which could cause situational visual impairment for users \cite{tigwell2018Situational}.

\subsubsection{Use Facial Recognition Instead Of Entering Pickup Code}

Facial recognition is a popular technology in the field of Machine Learning and Computer Vision. Previous studies on facial recognition technology within the BLV population mainly focused on how this technology could help BLV people recognize others \cite{Jin2021SmartcaneFace, Kianpisheh2019Face}. Facial recognition can be performed through the camera on the KuaiDiGui's machine to confirm the identity of the BLV recipient. The advantage of this approach is that BLV people no longer need to interact with the touch screen, they do not need to enter pickup codes, they can avoid encountering unexpected situations such as pop-up advertisements, and they can avoid interaction with the complex KuaiDiGui task flow. However, facial recognition technology has its limitations. First, facial recognition requires users to face the camera in the right direction. BLV people may have difficulty knowing whether the facial image captured by the camera meets recognition conditions and they will not know when the recognition process is completed. Courier companies will need to collect its user's facial information first in order to deploy this approach. This can be a privacy and security concern for users who do not want to share this information with the courier company.

\begin{figure*}[t]
  \centering
  \includegraphics[width=\linewidth]{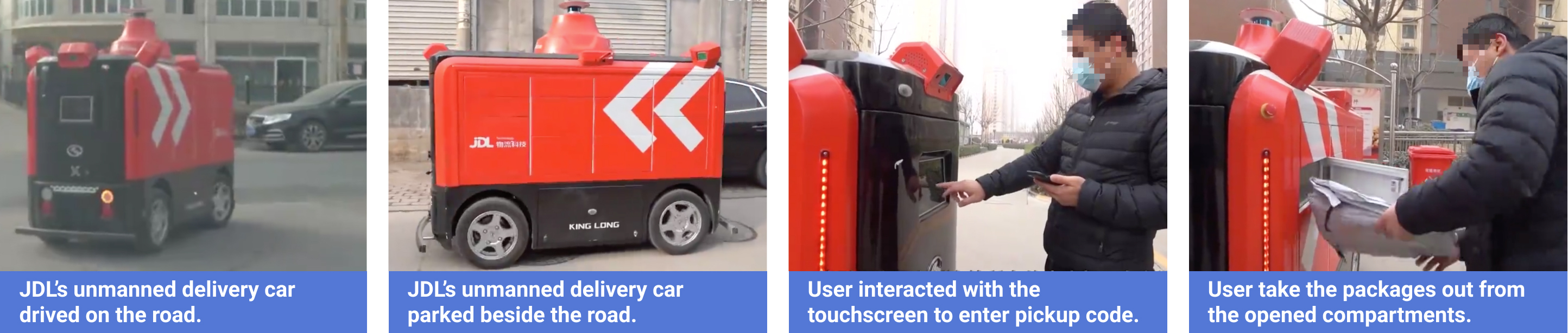}
  \caption{Jingdong unmanned auto-driving delivery car}
  \label{fig:JDcar}
  \Description{Four images that show the process of how JDL unmanned delivery car works, including unmanned delivery car driving on the road, unmanned delivery car parking beside the road, user’s interaction with the touchscreen of unmanned delivery car, and user taking out his package from an opened compartment.}
\end{figure*}

\subsection{How does the degree of vision impairment affect the package fetching process?}
\rv{We found that the degree of visual impairment has a significant impact on package fetching. Participants who are blind with residual vision or low vision can rely on their sight to complete some tasks more easily than totally blind participants. For example, low vision participants can see whether there is a package on the ground,  while totally blind participants might need to use their limbs and white canes to touch the area to feel for packages. Participants who had experience with using KuaiDiGuis all have residual vision. This is because they can utilize the interface by getting extremely close to them and read the interface little by little. Totally blind participants all failed to use KuaiDiGuis independently because of the lack of screen reader feature.}

\rv{Residual vision should be seriously factored when developing approaches to improve the accessibility of package fetching. Take the package localization approach as an example. Zhao et al. proposed an AR visualization technique that projects colored labels on stairs through a head-mounted AR to help visually impaired people navigate the stairs in their environment. This image processing technology can help to enhance the outline of packages during package localization. In this way, those with low vision can easily find packages in the environment, especially packages of a smaller size} \cite{yuhang2019AR}. \rv{Approaches that affect the interface of KuaiDiGuis (e.g. having larger icons on the interfaces, enhancing the color contrast of UI elements and background, and having magnifier feature) would also only work well if users have a certain level of residual vision. However, the AR approach cannot be applied to the totally blind population as it requires the users to have residual vision. Other modalities should be investigated for totally blind users. For example, an audio approach that voices directional navigation and feedback might benefit all types of vision impairment.}

\subsection{What are the impacts of automation and unmanned technology?}
In addition to the large-scale use of KuaiDiGuis in China, other unmanned delivery technologies (e.g. drone delivery, auto-driving car delivery) also started to merge in people's daily lives. With the advancement of machine learning and sensor technology, drones and unmanned vehicles began to appear in the package delivery process. Jingdong Logistics utilizes unmanned delivery vehicles and has obtained the official delivery vehicle license in Beijing in early 2021 \cite{JDnews}. As shown in Figure \ref{fig:JDcar}, to deliver a package with an unmanned vehicle, the delivery worker must first put the packages into the delivery vehicle's compartments and then the vehicle will automatically drive to the receiving address \cite{JDCar}. Users can interact with the unmanned vehicle via touchscreens and then take out their packages from the compartments in the vehicle \cite{JDCar}. The unmanned delivery vehicle is close to a self-driving KuaiDiGui. There are many compartments for storing packages, and there are also touch screens for people to touch and interact with the delivery vehicle. With this approach, the BLV people no longer need to travel long distances with an unfamiliar route to reach pickup sites. However, finding the parking location of the unmanned delivery vehicle could pose a potential challenge. Even with new technologies for express delivery, if the end user still needs to interact with a machine in the KuaiDiGui form, the BLV community will continue to encounter the same KuaiDiGui related problems when interacting with unmanned vehicle delivery service. 
%Using drone to perform delivery tasks is also a popular research area in the academic community \cite{Seo2016dronesecurity, Donkyu2018droneb, Park2016Drone,Milhouse2015drone}. Donkyu et. al. proposed a battery-aware delivery scheduling algorithm for delivery drones and demonstrated that with the same battery capacity, the new mode can carry more weighs compared to traditional one \cite{Donkyu2018droneb}. However, little previous drone delivery research focused on the accessibility issues during the delivery process. 
Therefore, engineers and designers of unmanned delivery products should take the needs of BLV people, like the ones uncovered in this study, into their design consideration.

\section{LIMITATION and future work}
To our knowledge, this is the first study focusing on BLV people's practices and challenges of package fetching experiences in light of the rapid change in package delivery service in recent years. Our study uncovers the needs and challenges that BLV users encounter during the package fetching process and thus provides a rich set of technical and social problems for HCI and accessibility researchers to address. However, it is worth noting that our BLV participants lived in a metropolitan area of a developed city in China. Although we firmly believe that many practices and challenges uncovered in this study are applicable to BLV communities in other parts of the country and other countries, cultural and infrastructural differences in different regions might shape their BLV people's practices and challenges. Thus, it is worth conducting similar studies in other countries and regions to uncover the similarities and differences across different BLV communities.  

Second, we only interviewed BLV people to gain their perspectives of package fetching experiences. However, there are multiple stakeholders if we aim to improve the user experience of the package fetching process. Future research should include other stakeholders, such as sighted people, delivery workers, and CEP companies, to better understand different perspectives to build an inclusive and accessible delivery service for all. 
We only briefly touched on participants' impressions on unmanned delivery technology (e.g., auto-driving vehicle delivery, drone delivery) that was recently emerging in the industry. Future work could further investigate BLV people's experience with such unmanned delivery technology and uncover directions for improvement.

\section{CONCLUSION}
We have presented a semi-structured interview study with 20 BLV people living in China to investigate their practices and challenges of the package fetching process and their attitudes towards novel delivery technologies. Our findings show that BLV participants tended to use their limbs or white canes to explore the environment and localize the target package from many similar packages of the residents in the area. When recognizing packages, BLV participants often identified the package label via touch or residual vision, and then use their smartphone to take photos to either zoom in to read or to use an OCR application to read out the recipient information. However, participants commonly needed to try many times until they captured a reasonably good photo to use. As a result, they tended to be reluctant to perform recognition by themselves. Moreover, our study also uncovered their practices and challenges with KuaiDiGuis, the self-service package pickup machine. we found that KuaiDiGuis lacked support for common accessibility features, such as screen readers, voice prompts, and physical keypads. Consequently, participants often failed to use KuaiDiGuis independently and needed to seek help from sighted people. Our findings also revealed how they sought help and managed their visual impairment identity. 
Based on these findings, we further discussed the potential design and research opportunities to improve the accessibility of the package fetching process. 
As a first work to investigate BLV people's practices and challenges of package fetching experiences, our work sheds light on what works and what does not work for BLV users and highlights many technical and social challenges for HCI and accessibility researchers to address.

\bibliographystyle{ACM-Reference-Format}
%\balance
\bibliography{main.bib}

% \section{Appendices}
% \label{appendices}

\end{document}